# Fruit harvesting: A potential threat to the persistence, spatial distribution, and establishment of plants

Mozzamil Mohammed[1,2,*], Åke Brännström[2,3], Pietro Landi[4], Ulf Dieckmann[2,5]

1. Institute for Chemistry and Biology of the Marine Environment, University of Oldenburg, Oldenburg, Germany.

2. Advancing Systems Analysis Program, International Institute for Applied Systems Analysis, Laxenburg, Austria.

3. Department of Mathematics and Mathematical Statistics, Umeå University, Umeå, Sweden.

4. Applied Mathematics Division, Department of Mathematical Sciences, Stellenbosch University, Stellenbosch, South Africa.

5. Complexity Science and Evolution Unit, Okinawa Institute of Science and Technology Graduate University (OIST), Onna, Okinawa 904-0495, Japan.

## Abstract

Plant-frugivore interactions play a central role for plant persistence and spatial distribution by promoting the long-range dispersal of seeds by frugivores. However, plant-frugivore interactions are increasingly being threatened by anthropogenic activities. An important anthropogenic threat that could expose plant-frugivore systems to extinction risk is fruit harvesting. Here, we develop an individual-based and a pair-approximation model of plant-frugivore-human interactions to elucidate the effects of human harvesting of fruits on plant establishment, persistence, and spatial distribution. Our results show that frugivores strongly affect global density of plants and gradually shift their spatial distribution from aggregated to random, depending on the attack rate and dispersal efficiency of frugivores. We find that, in the absence of frugivores, plants experiencing intense fruit harvesting cannot persist even if their fecundity is high. In the presence of frugivores, fruit harvesting profoundly affects the global dispersal of seeds and thus changes the spatial distributions of plants from random to aggregated, potentially causing plant extinction. Our results demonstrate that sufficiently efficient frugivores mitigate the negative impact of fruit harvesting on plant populations and enable plant establishment precluded by harvesting. Taken together, these results draw attention to previously underappreciated impacts of fruit harvesting in plant-frugivore-human interactions.

* Correspondence to: mozzamilm@gmail.com

# 1 Introduction

Frugivorous species, i.e., animals that feed on fruits, play a central role for plant communities by providing seed-dispersal services (Howe 1984, Keenan et al. 1997, Herrera 1989 & 2002, Bascompte & Jordano 2007, Jordano et al. 2011). Frugivorous animals gain nutritious food from fruit consumption while dispersing the seeds encapsulated in the fruit pulps. Plant-frugivore interactions benefit both partners and strongly affect plant persistence and spatial distribution (Mohammed et al. 2018). For instance, recent empirical work (Beckman et al. 2018) confirms that long-distance dispersal ability (often by efficient frugivores) is positively correlated with fast life-history strategies of plants favouring increased population sizes, in contrast to short-distance dispersal. However, frugivorous seed dispersal is sometimes risky for plants when the dispersal cost is high (Janzen 1970, 1971, Herrera 2002, Howe & Estabrook 1977). Inefficient frugivores can gradually drive plants to extinction by seed predation. Plants with low local dispersal ability (i.e., the ability to disperse seeds in the neighbourhood of focal plants) cannot persist if interacting with inefficient frugivores (Mohammed et al. 2018).

The mutualistic interactions between plants and their frugivores are increasingly being threatened by anthropogenic activities (Markl et al. 2012). Anthropogenic threats have left, over long periods of time, negative impacts on global diversity and ecosystem services and have significantly shaped our presently experienced ecological conditions (Mishra et al. 2004, Hunter 2007, McConkey et al. 2011, Diaz et al. 2016). The loss of frugivore species could profoundly affect the abundances of economically and socially important plants (Egerer et al. 2018), and it has been suggested that this could even lead to plant extinctions (Caughlin et al. 2015, Perez-Mendez et al. 2015). Therefore, the study of the impact of anthropogenic disturbances on plant-frugivore systems is of central importance for nature conservation.

An important example of an anthropogenic disturbance that can expose plant-frugivore systems to extinction risk is fruit harvesting. Fruit harvesting can affect the abundances of plants and their frugivores, as well as frugivore fruit-consumption rates and thus the long-range dispersal of seeds (Moegenburg & Levey 2003, Markl et al. 2012, Edeline 2016). Harvesting often has disadvantages (Law & Salick 2005) for the demography and fitness-related phenotypic traits of affected populations, depending on its intensity (Bauer et al. 2013). On the one hand, fruit harvesting increases the mortality of seeds; hence, exceedingly intense fruit harvesting can lead plants and their specialists frugivores (relying entirely on plants for food) to extinction. On the other hand, the chances of plant persistence decrease with an increased harvesting rate of fruits. Plant densities increase with an increased long-range dispersal of seeds by efficient frugivores (Mozzamil et al. 2018). Recent studies (Kuparinen & Festa-Bianchet 2017) suggest that harvesting causes reductions in the resilience and recovery capacity of the affected populations, and it is therefore important to understand how plant communities and their frugivores respond to human harvesting of fruits.

While the impact of human harvesting of fruits on plant persistence is widely acknowledged, its impact on the spatial distributions of plant communities and plant establishment remains an entirely open question. Here we develop and analyze a spatially explicit individual-based model of plant-frugivore interactions to investigate the ecological response of plant-frugivore systems to human harvesting of fruits. In particular, we investigate the effects of fruit harvesting on the persistence and spatial distributions of plants, and elucidate the ecologically important role played by frugivores for mitigating the potential negative impacts of fruit harvesting on plant populations. We consider key characteristics of frugivorous seed dispersal, including seed-reproduction rates, fruit-consumption rates, dispersal efficiencies of frugivores, and germination probabilities of seeds. We then account for the impacts of fruit harvesting on plant-frugivore systems and investigate the resulting effects on the spatiotemporal dynamics of plants. Finally, we use pair-approximation techniques to derive mathematical formulas describing parameter-dependent conditions under which plant establishment is possible. Our results provide useful ecological insights for the management of plant-frugivore-human interactions and for the conservation of plant and animal communities experiencing the pressure of anthropogenic disturbances.

## 2 Methods

### 2.1 Individual-based model of plant-frugivore-human interactions

We develop a spatially explicit individual-based model of plant-frugivore mutualistic interactions through seed dispersal to investigate the effects of anthropogenic fruit harvesting on plant and animal densities and the spatial distribution and establishment of plants. We consider a finite regular lattice on which a seed is either dispersed locally by the parent plant to neighbouring lattice sites or dispersed globally by frugivorous animals to anywhere on the lattice (Harada & Iwasa 1994, Mohammed et al. 2018). Each lattice site is either empty or occupied by an individual plant (Fig. 1). Plants produce seeds at rate $r_P$ and naturally die at rate $d_P$. Without frugivores, plants can only disperse seeds in their neighbourhood or die naturally. Considering frugivorous seed dispersal, we divide seeds into two groups: seeds eaten by animals are dispersed globally on the lattice and seeds dispersed by parental plants are dispersed locally in the parental plants' neighbourhood. For simplicity, we assume that each fruit contains a single seed. Animals are assumed to have a random spatial distribution and to exhibit random foraging behaviour, according to which an individual frugivore encounters a seed with attack rate $a$. The attack rate of all frugivores per seed is thus given by $aA$, where $A$ is the animal density. By using a Holling type-II functional response, the effective frugivore attack rate per seed is given by $c = aA/(1 + \tau r_P P)$, where $\tau$ is the handling time (i.e., the time frugivores spend to consume fruits) and $P$ is the global plant density.

We denote by $s$ the intrinsic rate at which non-eaten seeds are dispersed naturally by parental plants in their own neighbourhoods and by $c$ the rate at which seeds are eaten by frugivores. The proportion of seeds that are naturally dispersed by plants thus is $\frac{s}{c+s}$, and the proportion of seeds that are dispersed by frugivores is $\frac{c}{c+s}$. Before seeds are successfully dispersed by frugivores, the eaten seeds experience a risk while passing through the digestive system of frugivores, so the proportion of seeds that are dispersed globally by frugivores is $\mu \frac{c}{c+s}$, where μ measures the frugivore dispersal efficiency, and conversely, $(1 - \mu)$ measures the dispersal cost. Locally and globally dispersed seeds must fall on empty lattice sites before they can germinate with probability $g$; otherwise, they die. The local and global lattice sites on which the dispersed seeds fall are chosen at random.

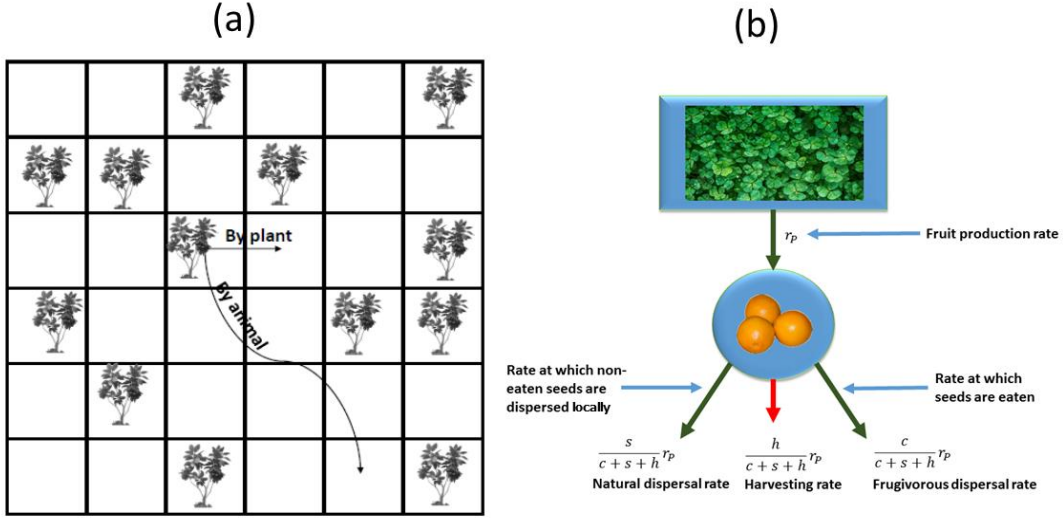

**Fig. 1:** a, Schematic representation of local seed dispersal by plants and of global seed dispersal by animals in the lattice-based model. Each lattice site is either empty or occupied by an individual plant. b, Flow chart illustrating the modelling of frugivorous seed dispersal with fruit harvesting.

We now introduce the effects of fruit harvesting on the plant-frugivore system by assuming that some fruits are being harvested at rate $h$. The proportion of seeds that are dispersed locally by plants then equals $\frac{s}{h+c+s}$, the proportion of seeds that are dispersed globally by frugivores then equals $\mu \frac{c}{h+c+s}$, and the proportion of seeds that are lost to harvesting equals $\frac{h}{h+c+s}$.

The birth of a new plant can thus take place in two ways: birth due to the local dispersal of seeds by parental plants in their neighbourhood and birth due to the global dispersal of seeds by frugivores across the entire lattice. Also, the death of a plant can take place in two ways: natural mortality of plants with rate $d_P$, and harvesting-induced mortality (see below). Frugivores are assumed to be generalists (i.e., they have alternative food resources other that plants) and they grow according to the logistic equation with population growth rate $r_A A(1 - A/K)$, where $r_A$ is the frugivore intrinsic growth rate and $K$ is their carrying capacity. Plants contribute to the population of frugivores at rate proportional to the product of the global plant density $P$, fruit production rate $r_P$, fruit consumption rate (above), and the efficiency of frugivores to convert the eaten fruits to their population $\alpha$. The birth rate of frugivores due to fruit consumption is therefore given by $\alpha \frac{c}{c+s+h} r_P P$. In our individual-based model, all these six birth-death events of plants and frugivores occur at random.

## 2.3 Pair-approximation model of plant-frugivore-human interactions

Using the pair-approximation method (Harada and Iwasa 1994) coupled with the mean-field-approximation, we approximate the behaviour of our individual-based model described above. Pair approximation is a method of constructing an ordinary differential equation model of populations dynamics and considers both the local and global densities of populations. Pair approximation has been employed in ecology to describe plant dynamics and the interaction between plant and animals (Liao et al. 2013; Mohammed et al. 2018). It correctly captures the average dynamics of plants predicted by individual-based models (Harada and Iwasa 1994; Ellner 2001). Here, we will use the pair-approximation to derive ordinary differential equations describing the dynamics of the global and local densities of plants and describe the dynamics of frugivores using the mean-field approximation. A brief introduction about the pair-approximation modelling techniques is provided as in *Appendix A*.

The global plant density is defined as the frequency of plants; that is, the total number of plants divided by the total number of potential lattice sites, and denoted as $P_+$. Note that pair approximation considers infinite lattice size (Harada and Iwasa 1994). In the pair-approximation method, the global plant density, $0 \leq P_+ \leq 1$, represents the probability that a randomly chosen habitat site is occupied by a plant individual. The probability that a randomly chosen global site is empty is thus given as $P_0 = 1 - P_+$. This probability is crucial for determining the successful dispersal of seeds dispersed globally by frugivores.

The local plant density is defined as the conditional probability that a randomly chosen site (locally in the neighbourhood of a focal plant) is occupied, and it is denoted as $q_{+|+}$. Mathematically, the local plant density is defined as $q_{+|+} = P_{++}/P_+$, where $P_{++}$ is the joint probability that a randomly chosen pair of sites are both occupied. By definition $q_{+|+}$ cannot be zero (*i.e.*, $0 < q_{+|+} \leq 1$) as the global plant density, $P_+$, approaches zero. The conditional probability that a randomly chosen nearest-neighbouring site of an occupied site is empty is given as $q_{0|+} = 1 - q_{+|+}$. This conditional probability of finding an empty local site of a focal plant will be required for the successful germination of locally dispersed seeds. The local plant density characterises the spatial clustering of plants relative to the global plant density. That is, plants are segregated as the local density is less than the global density ($q_{+|+} < P_+$), aggregated or clustered as the local density is greater than the global density ($q_{+|+} > P_+$), and randomly distributed as the local and global densities are equal ($q_{+|+} = P_+$). In the latter case, the pair-approximation method approaches the mean-field approximation.

We now use the pair-approximation modelling technique to approximate the behaviour of the individual-based model of plant-frugivore-human interactions. In addition to readily yielding an explicit description of the local density of plants, pair approximation allows a faster way of analyzing the model and of investigating the effects of model parameters on its dynamics.

The dynamics of the global density of plants is described as

$$\dot{P}_+ = -d_P P_+ + g r_P (1 - q_{+|+}) \frac{s}{c+s+h} P_+ + \mu g r_P (1 - P_+) \frac{c}{c+s+h} P_+ - \frac{h}{c+s+h} r_P P_+, \quad (1)$$

The first term represents the natural death of plants, the second term represents the birth of new plants due to the local dispersal of seeds by parental plants in their own neighbourhood, the third term represents the birth of new plants due to the global dispersal of seeds by frugivores, and the last term represents the death of plants due to fruit harvesting. The probabilities of finding an empty local site and empty global site for potential local and global germination of seeds are, respectively, given by $(1 - q_{+|+})$ and $(1 - P_+)$.

As in the description of our individual-based model, we assume that animals are generalists and utilize food resources other than the focal plant species. Therefore, in the absence of plants, they grow according to logistic dynamics. Considering the plant-frugivore mutualistic interactions, the pulp of the eaten fruits is converted into new frugivores with a conversion efficiency $\alpha$.

Thus, the dynamics of the global density of frugivores is described as

$$\dot{A} = r_A A \left(1 - \frac{A}{K}\right) + \alpha \frac{c}{c+s+h} r_P P_+, \qquad (2)$$

where $A$, $r_A$, and $K_A$ are the global density, intrinsic growth rate, and carrying capacity of frugivores, respectively. The first term in Eq. 2 refers to the logistic growth of frugivores, while the second term describes the positive contribution of plants to frugivore growth. Alternatively, frugivores could be assumed to be specialists, i.e., obligatory mutualists that completely depend on the fruit pulps of the focal plant species for food (Donoso et al. 2017). In such a case, the first term in Eq. 5 would be replaced by $-d_A A$, where $d_A$ is the natural death rate of animals. Since generalist and specialist frugivores have been shown in an earlier study to have the same effects on plant dynamics and plant spatial distributions (Mohammed et al. 2018), here we mostly focus on examining generalist frugivores. Obviously, specialist frugivores die whenever the focal plants go extinct, whereas specialist frugivores converge to their carrying capacity in the absence of the focal plant species.

The dynamics of the local density of plants is describe as

$$\dot{q}_{++} = -\left(d_P + \frac{h r_P}{s+c+h}\right) q_{++} + \frac{g r_P}{z} \frac{s}{s+c+h} \left(2 - z\, q_{+|+} + 2(z-1)\frac{(1 - q_{+|+}) P_+}{(1 - P_+)}\right)(1 - q_{+|+}) + g\mu r_P \frac{c}{s+c+h}\left(2 P_+ - q_{+|+} - P_+ q_{++}\right). \qquad (3)$$

Eqs. 1-3 provide a system of three nonlinear ordinary differential equations governing the ecological dynamics of plant-frugivore-human interactions. The derivation of Eq. 3 is provided in *Appendix B*.

**Table 1.** Variables and parameters used in our frugivorous seed-dispersal model.

| Symbol | Description | Reference value | Range |
|---|---|---|---|
| $P_+$ | Global plant density | | [0, 1] |
| $q_{++}$ | Local plant density | | [0, 1] |
| $A$ | Frugivore density | | [0, ∞) |
| $d_P$ | Natural death rate of plants | 0.6 | [0, ∞) |
| $r_P$ | Seed-production rate | 5 | [0, ∞) |
| $s$ | Natural dispersal rate | 10 | [0, ∞) |
| $g$ | Germination probability per seed | 0.3 | [0, 1] |
| $h$ | Fruit-harvesting rate | 5 | [0, ∞) |
| $r_A$ | Intrinsic growth rate of frugivores | 0.5 | [0, ∞) |
| $K$ | Carrying capacity of frugivores | 100 | [0, ∞) |
| $z$ | Number of nearest-neighbouring sites | 4 | [0, ∞) |
| $N$ | Total number of lattice sites | 2500 | |
| $a$ | Frugivore (per-capita) attack rate | 0.8 | [0, ∞) |
| $\mu$ | Dispersal efficiency | 0.8 | [0, 1] |
| $\alpha$ | Conversion rate | 0.1 | [0, 1] |
| $\tau$ | Handling time | 0.0001 | [0, ∞) |

## 2.3 Numerical analysis

### Algorithm for individual-based model

We use the event-driven Gillespie stochastic simulation algorithm (Gillespie 1977) to implement the individual-based model. This allows the generation of statistically correct trajectories of the stochastic version of first-order differential equations of a given system (Gillespie 1977). The Gillespie algorithm is based on the observation that each event is chosen randomly to occur according to its event rate and that the waiting time for the next event to occur is exponentially distributed according to the total event rate. By implementing the Gillespie algorithm, we can simulate the stochastic birth-death process of mutualistic plants and frugivores under fruit harvesting.

We summarize below the steps of the algorithm to run the simulation of the model:

1. Initialize populations and time.
2. Compute the rates (a-f) at which each type of event $E_i$ occurs.

   a. Plant-independent animal growth rate: $r_A A(1 - \frac{A}{K})$

   Animals are assumed to grow according to the logistic equation in the absence of plants. When a random natural animal birth event is chosen (see below), then the abundance of animals is increased by one.

   b. Animal birth rate due to fruit consumption: $\alpha \frac{c}{c+s+h} r_P P$

   New animal biomass is formed when a fruit is successfully consumed by an animal, and the conversion efficiency of animals is described by α. When a random fruit-consumption event is chosen (see below), then the abundance of animals is increased by one.

   c. Plant natural death rate: $d_P P$

   Plants die naturally. When a random natural plant death event is chosen (see below), then a randomly selected plant individual is removed.

   d. Plant death due to harvesting: $\frac{h}{c+s+h} r_P P$

   The harvested seeds are counted as harvesting-dependent mortality of plants. When a random harvesting event is chosen (see below), then a randomly selected plant individual is removed.

   e. Plant birth rate due to local dispersal by plants: $g \frac{s}{c+s+h} r_P P$

   Plants disperse seeds locally in their own neighborhood without support from animals. When a random local dispersal event is chosen (see below), then a random site is selected from the neighborhood of a randomly chosen occupied location. If empty, a new individual plant is placed.

   f. Plant birth rate due to global dispersal by frugivores: $\mu g \frac{c}{c+s+h} r_P P$

   Animals encounter plants, consume the fruits, and disperse the seeds encapsulated in the fruits with some probability of dispersal success. When a random animal dispersal event occurs, a random site is selected anywhere in the landscape. If empty, a new individual plant is placed.

3. Draw two random numbers, $R_1$ and $R_2$, uniformly distributed between 0 and 1.
4. Choose the next time step using $R_1$, with a waiting time of $-\ln(R_1) / \sum_i E_i$ with $E_i$ for $i = a, b, \ldots, f$ ranging over all six event rates given above.
5. Update time by adding the chosen waiting time.

6. Choose the next event using $R_2$, with each event type $j = a, b, ..., f$ having a probability of $E_j / \sum_i E_i$ to be chosen.
7. Update population numbers by realizing the chosen event as described above under the rates (a-f).
8. Repeat until the maximum time is reached.

*Local plant density, statistical equilibrium, and degree of plant clustering*

The local plant density, $q_{+|+}$, is well-defined in the pair-approximation model (see Eqn. 3 and its derivation), but in the individual-based model we compute it as follows. We consider four neighbouring sites (direct neighbours) of focal plants (defined as $z$ in the pair-approximation model) to describe the local plant dynamics. Every unit of the simulation time, we count the number of occupied sites of these four neighbouring sites for each focal plant. We then take the average of the locally occupied sites and obtain a mean value for the local plant density each unit of time.

To compute the stationary equilibrium abundances of plants from the individual-based model, we waited until 75% of the total simulation time (100 units of time) has elapsed and then averaged over the last 25% of the time to avoid transient effects in our estimations. We found almost perfect matching between the equilibria obtained from the individual-based model and the pair-approximation model. We have tested variability among different simulation runs and found very small variability (*Appendix C*).

The degree of plant clustering, using to assess plants' spatial distribution, is defined as the ratio of the local plant density and the global plant density. We further used the common logarithmic scale, i.e., $\log(q_{+|+}/P_+)$.

The numerical simulation of both the individual-based model and pair-approximation model is implemented in MATLAB. Parameters used to simulate the plant-frugivore-human system are provided in *Table 1*.

## 3 Results

We investigate the effects of frugivore animals and human harvesting of fruits on the persistence and spatial distribution of plant populations, and on plant establishment. Using the individual-based and pair-approximation models, we focus on numerically analysing the influences of the attack rate and dispersal efficiency of frugivores and the fruit harvesting rate on the local and global plant densities and plant spatial distributions, since they are the main parameters describing the interactions and driving the dynamics. We then perform further analysis for the dependence of the other model parameters on the plant-frugivore-human system and provide them as supplementary materials. We finally derive analytical establishment conditions of plants using the pair-approximation model.

## 3.1 *Frugivores typically increase the global density of plants when highly efficient*

We first analysed the plant-frugivore system without human harvesting of fruits, and investigated the effect of frugivore animals on the local and global plant densities and on the spatial distribution of plants (*section* 3.2). Our results agree with previous findings (Mohammed et al. 2018) and demonstrate that frugivores enable the persistence of plant populations that would not otherwise persist, provided that the global dispersal cost is low (Fig. 2a,c, 4a,c and 5a). However, the effect of frugivores on global plant density depends strongly on the efficiency and the attack rate of frugivores (Fig. 5a,c and 8a). The efficiency of frugivores is quantified in the model by the fraction of seeds ($\mu$) eaten by frugivores that will be safely passing through the animal digestive system and deposited successfully. We found that efficient seed-dispersers (large $\mu$) increase global plant density through global dispersal of seeds while inefficient dispersers (small $\mu$) behave as seed predators and drive plants to extinction (Fig. 5a and 8a). The global equilibrium abundance of plant increases with the efficiency of frugivores. Our results exhibit a transcritical bifurcation point on the dispersal efficiency axis ($\mu$), splitting the curve of the global equilibrium abundance of plants onto stable extinction (driven by inefficient frugivores) and stable persistence (driven by efficient frugivores).

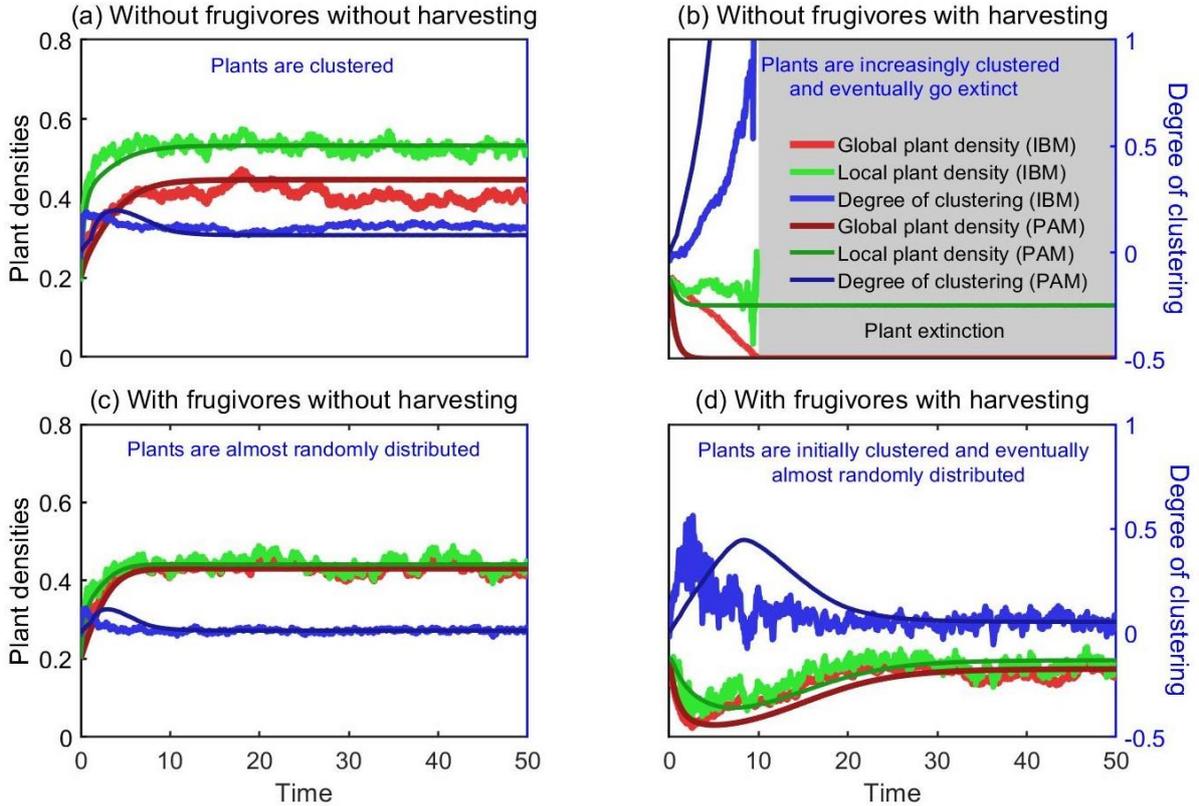

**Fig. 2**: **Frugivores typically increase the global density of plants when highly efficient**. Frugivores allow the persistence of plants through long-distance dispersal of plants seeds while human harvesting of fruits drive the extinction of plant populations in the absence of frugivores and reduces plant global density in the presence of frugivores. Panels (a) – (d) show trajectories of the individual-based model (IBM) (thick bright curves) and the pair-approximation model (PAM) (thin dark curves). The red curves correspond to the global plant density, the green curves correspond to the local plant density, and blue curves correspond to the degree of plant clustering. Changes in plant density (left y-axis) and degree of plant clustering (right y-axis) over time (x-axis) are shown for four different ecological scenarios: (a) without frugivores without fruit harvesting, (b) without frugivores with fruit harvesting, (c) with frugivores without fruit harvesting, and (d) with frugivores with fruit harvesting. The grey-shaded regions correspond to the extinction of plants. Parameters: $n = 50$, $d_P = 0.6$, $r_p = 5$, $s = 10$, $z = 4$, $g = 0.3$, $h = 5$, $\tau = 0.0001$, $a = 0.8$, $r_A = 0.5$, $\alpha = 0.1$, and $\mu = 0.8$.

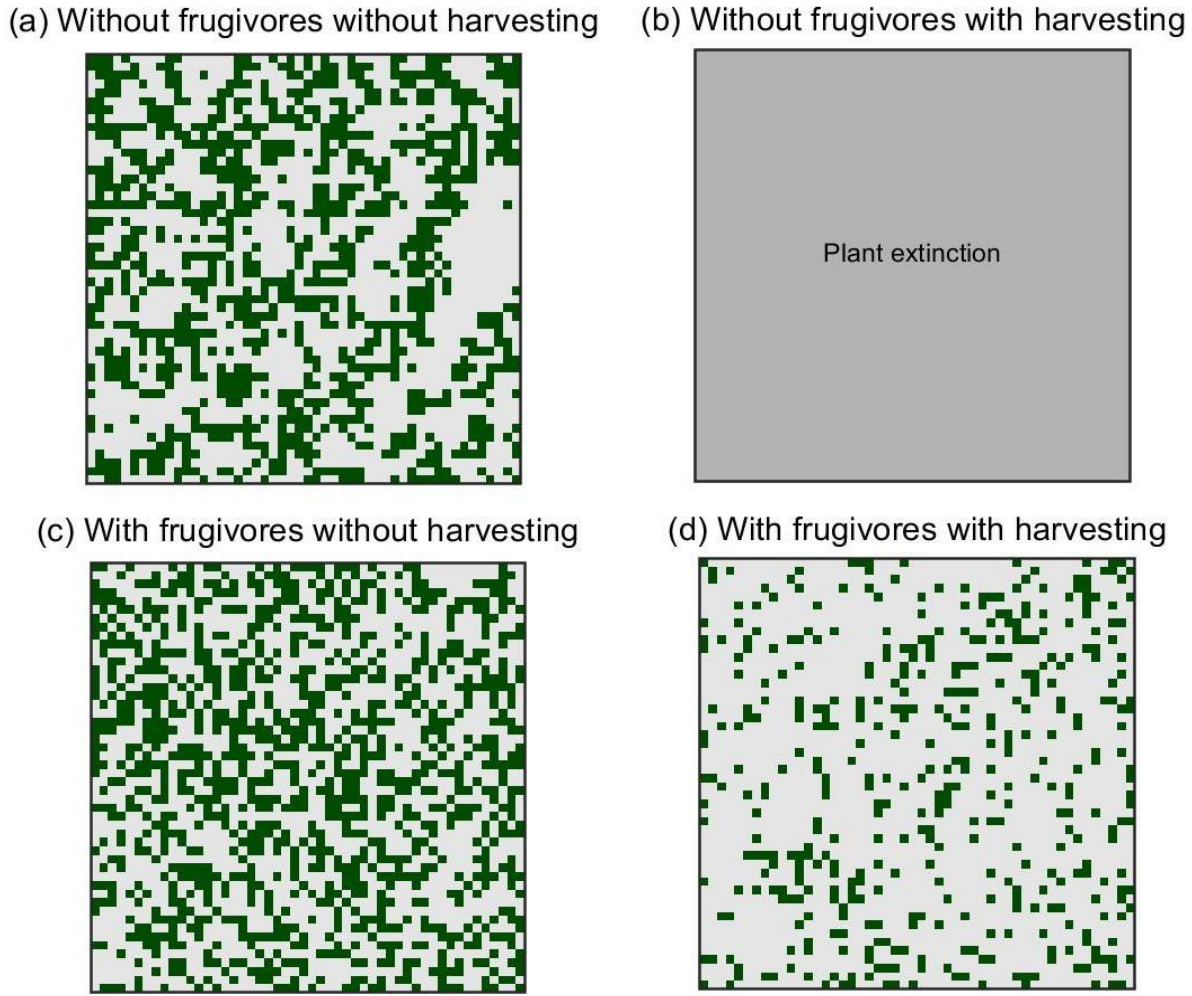

**Fig. 3**: **Frugivores make the spatial distribution of plants less aggregated**. Frugivores change plant spatial distribution from aggregated to random through the long-distance dispersal of plant seeds while human harvesting of fruits drives plant extinction in the absence of frugivores and reduces plant global density in the presence of frugivores. The x- and the y-axis represent the horizontal and vertical spatial locations, respectively. Four different ecological scenarios of the plant spatial distribution are shown from the individual-based model (a) without frugivores without fruit harvesting, (b) without frugivores with fruit harvesting, (c) with frugivores without fruit harvesting, and (d) with frugivores with fruit harvesting. The light grey-shaded regions correspond to empty locations, while the dark, grey-shaded region represents the extinction of plants. All parameters are as indicated in the caption of figure 2.

### 3.2 *Frugivores make the spatial distribution of plants less aggregated*

Here we focus on the effect of frugivore animals on the spatial distribution of plants without human harvesting of fruits. In the absence of frugivores, plants are only able to disperse their seeds in their neighbourhood, causing plant clustering (Fig. 2a and Fig. 3a). In line with previous findings (Mohammed et al. 2018), our results show that frugivores, by dispersing seeds globally over the habitat, change the spatial distribution of plants from aggregated to random (Fig. 2c and Fig. 3c), thus reducing local competition in the neighbourhood of focal plants. However, the effect of frugivores on plants' spatial distribution depends strongly on the characteristics of frugivores such as the attack rates and the dispersal efficiency (Fig. 5a,c). That is, efficient frugivores would have the potential of making the spatial distribution of plants completely random in space while inefficient frugivores would not be able to change the aggregated distribution of plants (Fig. 5a). The same applies to the frugivore attack rate (Fig. 5c).

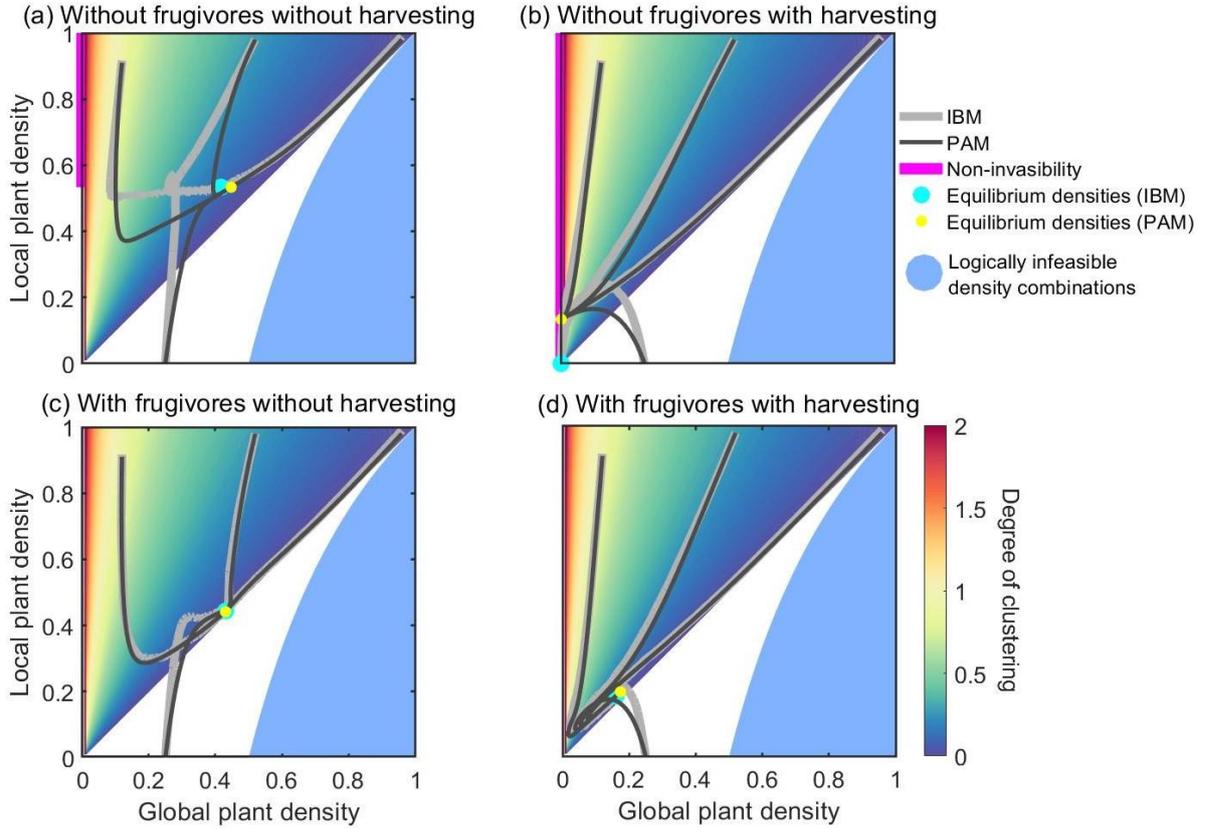

**Fig. 4**: **Fruit harvesting decreases the global density of plants and can drive them to extinction**. Frugivores change plant spatial distribution from aggregated to random through the long-distance dispersal of plant seeds while human harvesting of fruits drives plant extinction in the absence of frugivores and reduces plant global density in the presence of frugivores. Panels (a) – (d) show the joint dynamics of global plant density $P_+$ and local plant density $q_{+|+}$ from the individual-based model (grey curves) and the pair-approximation model (black curves) for four different ecological scenarios: (a) without frugivores and without fruit harvesting, (b) without frugivores and with fruit harvesting (c) with frugivores and without fruit harvesting, and (d) with frugivores and with fruit harvesting. Trajectories from the individual-based model are averaged over 50 stochastic realizations. The diagonal of each panel corresponds to the mean-field approximation, in which the global and local plant densities are assumed to be equal as plants are assumed to be randomly distributed in space. Combinations of the two densities above the diagonal (color-coded region; the colorbar is shown on the right of panel (d)) correspond to aggregated (clustered) plant distributions, whereas those below the diagonal (white regions) correspond to segregated (overdispersed) plant distributions. Filled cyan and yellow circles indicate stable equilibria from the individual-based model and the pair-approximation model, respectively. The light, blue-shaded regions (given by $2 - \frac{1}{P_+} \leq q_{+|+}$; *Appendix A*) are mathematically infeasible. The magenta vertical lines (given by Eq. 4) represent the local plant density at which plants can not establish from low abundances. All parameters are as indicated in the caption of Figure 2.

### 3.3 *Fruit harvesting decreases the global density of plants and can drive them to extinction*

We now focus on human harvesting of fruits and its effect on the abundance of plants and on their spatial distribution (*section* 3.4). Without frugivores, we found that intense fruit harvesting drives the extinction of plants (Fig. 2b, 3b and 4b). In the presence of frugivores, harvesting of fruits decreases the global plant density by causing decreased recruitment of plants through removal of fruits (Fig. 2d, 3d and 4d) and decreases the abundances of their frugivores (*Appendix C*). Our results demonstrate that high rates of fruit harvesting cause plant extinction even if seeds are efficiently dispersed by frugivores (*Appendix C*). We found that the degree at which the global plant density is decreased by harvesting depends largely on the intensity of fruit harvesting and the frugivores attack rate and dispersal efficiency (*section* 3.5).

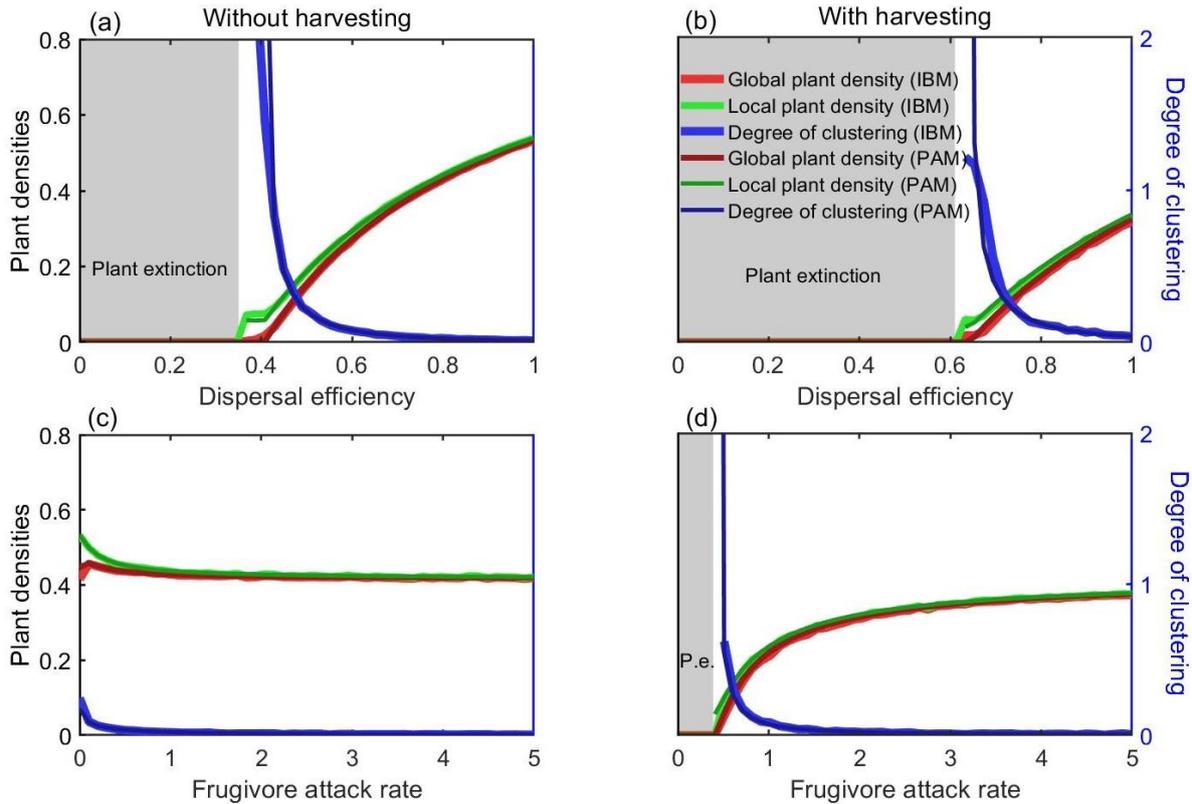

**Fig. 5: Fruit harvesting makes the spatial distribution of plants more aggregated.** Changes in the equilibrium of global plant density $P_+$ (left y-axis; red and green curves) and the degree of plant clustering (right y-axis; blue curves) as a function of the dispersal efficiency (panels a and b) and the frugivore attack rate (panels c and d) predicted by the pair-approximation and the individual-based model. The statistical equilibria of the individual-based model are the mean of the time-dependent trajectories and averaged over 50 stochastic realizations. The grey-shaded regions correspond to plant extinction. The first column and the second column represent changes in the equilibrium frequencies and clustering of plants without and with fruit harvesting, respectively. All other parameter values are as indicated in the caption of Figure 2.

### 3.4 *Fruit harvesting makes the spatial distribution of plants more aggregated*

Our results demonstrate that fruit harvesting affects the important role played by frugivores for the spatial distribution of plants by decreasing the global dispersal rates of seeds (Fig. 5). While frugivores change the spatial distribution of plants from aggregated to random, we found that fruit harvesting changes the spatial distribution from random to aggregated (Fig. 5). We found strong dependence of the degree of plant clustering on the intensity of fruit harvesting. That is, high harvesting rates leads to high degree of plants clustering (Fig. 6). Depending on its intensity and on the global dispersal rate of seeds, plant clustering caused by fruit harvesting can potentially drive plants to extinction as competition in the neighbourhood of focal plants is higher and the plants are kept at lower global abundances. However, the effect of fruit harvesting on degree of plant clustering depends on the attack rate and dispersal efficiency of frugivores (*section* 3.5).

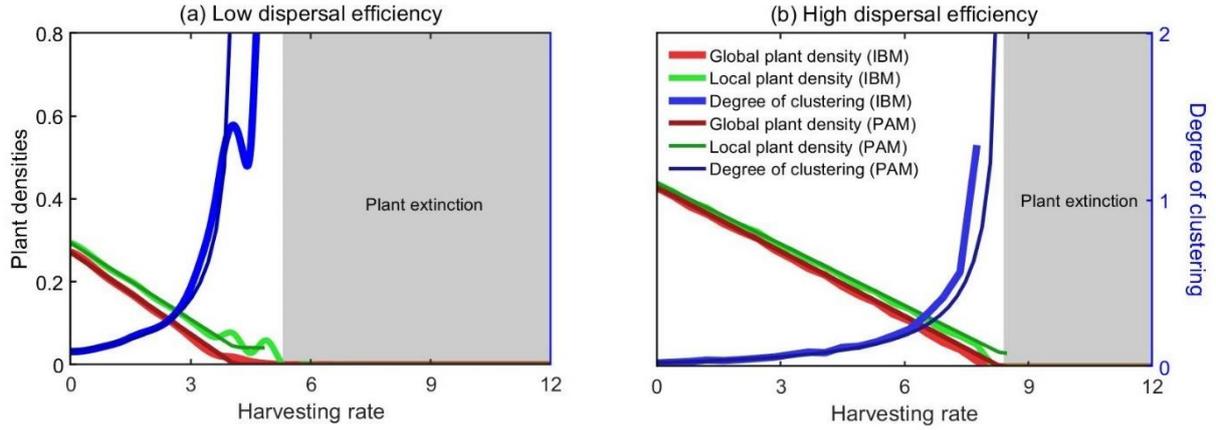

**Fig. 6: Frugivores mitigate the negative impact of fruit harvesting on plant persistence.** Changes in the equilibrium frequencies of global plant densities $P_+$ (left y-axis; red and green curves) and the degree of plant clustering (right y-axis; blue curves) as a function of fruit-harvesting rate predicted by the pair-approximation and the individual-based model. The statistical equilibria of the individual-based model are the mean of the time-dependent trajectories and averaged over 50 stochastic realizations. The grey-shaded regions correspond to plant extinction. The first column and the second column represent changes in the equilibrium frequencies and clustering of plants low and high dispersal efficiencies of frugivores, respectively. All other parameter values are as indicated in the caption of Figure 2.

## 3.5 Frugivores mitigate the negative impact of fruit harvesting on plant persistence

In the absence of frugivores, plants are only able to disperse their seeds in their own neighbourhoods at the cost of strong local competition and additional mortality caused by fruit harvesting. Thus, plants face a strong extinction risk. However, the negative influence of fruit harvesting on plant population can be mitigated by frugivore animals, depending on their dispersal efficiency and their carrying capacity in the absence of plants (Fig. 6 and 7). In the presence of efficient frugivores that disperse plant seeds at very low cost, fruit harvesting would have unpronounced impact on the persistence and spatial distribution of plants rendered by frugivores (Fig. 5a,b). The same applies in the presence of highly aggressive frugivores (Fig. 5c,d). When the frugivore attack rate is low, however, then fruit harvesting would have significant effect on the degree of plant clustering, even if frugivores are efficient seed-dispersers (Fig. 5d). In such a case, very small number of fruits are eaten by frugivores; thus, frugivores have a little impact on plants. Fruit harvesting also causes plant clustering but the degree of clustering caused by fruit harvesting is reduced by efficient seed dispersal by frugivores, especially those with high attack rates. We show the ecologically important role played by aggressive and efficient frugivores in mitigating the extinction risk experienced by plants from fruit harvesting through global dispersal of plant seeds.

We assumed that animals are generalists (i.e., they have food resources other than plants), and in the absence of plants, they grow until they reach their carrying capacity (see Eqn. 5). In our model set up, more external support of frugivores by the environment indicates that more animals will interact with each plant with some probability. Our results demonstrate that the carrying capacity of frugivores reduces the negative impact of fruit harvesting on plant global densities (Fig. 7). Under fruit harvesting, the global plant density increases with an increased external support of frugivores by the environment. Exceedingly intensive harvesting rates of fruits will drive plants to extinction even if the frugivores carrying capacity is high.

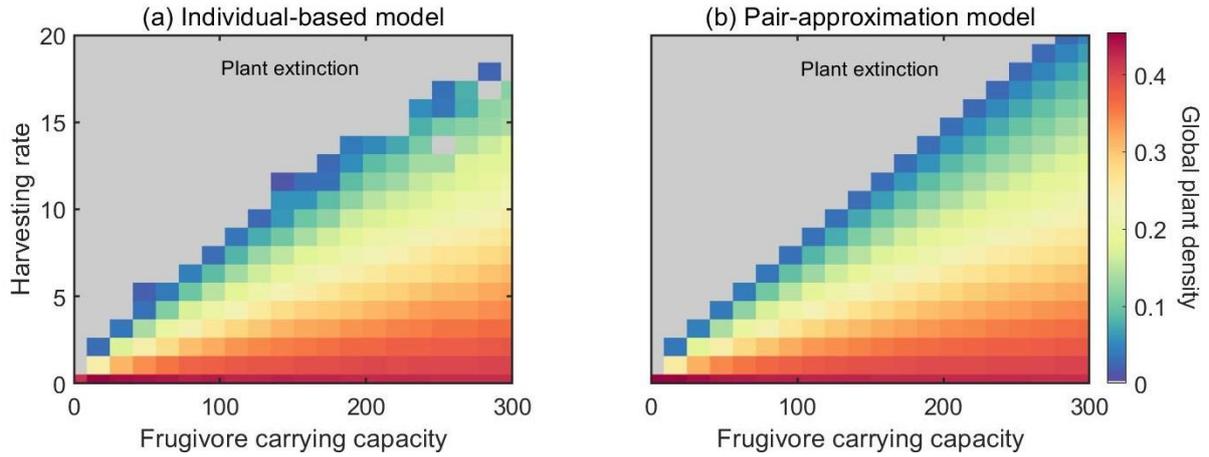

**Fig. 7: Frugivores mitigate the negative impact of fruit harvesting on plant persistence**. Changes in the equilibrium of global plant density $P_+$ (color-coded) driven jointly by frugivore carrying capacity (x-axis) and human harvesting of fruits (y-axis); predicted by the individual-based model (a) and the pair-approximation model (b). The grey-shaded regions correspond to the extinction of plants. All other parameter values are as indicated in the caption of Figure 2.

Our results have demonstrated the interplay between frugivore attack rate and dispersal efficiency in mitigating the negative impact of fruit harvesting on plants (Fig. 8). In particular, we found that highly efficient frugivores must necessarily have high attack rates to help plants persist under the pressure of fruit harvesting (Fig. 8b). In the presence of fruit harvesting, the global plant density increases as both the frugivore dispersal efficiency and the attack rate increase.

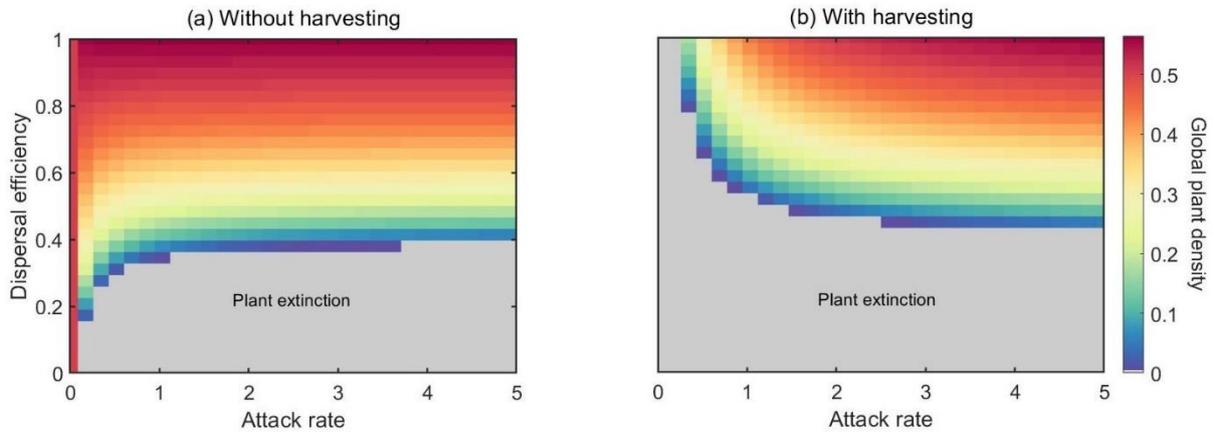

**Fig. 8:** Changes in the equilibrium of global plant density $P_+$ (color-coded) driven jointly by frugivore attack rate (x-axis) and dispersal efficiency (y-axis) without harvesting (a) and with harvesting (b). The grey-shaded regions correspond to the extinction of plants. All other parameter values are as indicated in the caption of Figure 2.

### 3.6 *Frugivores enable plants establishment otherwise precluded by fruit harvesting*

Using the pair-approximation model, we derive analytical conditions under which plants can stablish from low abundances or under which establishment is not possible. We analytically investigate the sensitivity of the plant equilibrium to all model parameters. To analyse plant establishment, we assume that animals are at their carrying capacity ($A = K$) while the global plant density, $P_+$, is set to zero. However, the local plant density $q_{+|+}$, defined by pair approximation as $P_{++}/P_+$, has a non-zero value even when the global plant density is zero.

By substituting $P_+ = 0$ and $A = K$ in Eqns. 4 and 5 and setting $\dot{P}_+ > 0$ we obtain a positive population growth rate of global plant density satisfying the inequality

$$q^*_{+|+} < 1 + \left(\frac{1}{grs}\right)[\mu graK - d_P(aK + s + h) - hr]. \tag{4}$$

Plants can establish from low abundance only if this inequality holds. Our analysis show that increases in the frugivores parameters ($\mu, a, K$) favour plant establishment while increases in the fruit harvesting rate ($h$) hinders the establishment. We numerically show the following: 1) without frugivores without harvesting plants can invade if the local plant density is below 0.5 (see magenta lines in Fig. 4), 2) without frugivores with harvesting plants can not invade even if plants are completely segregated in space, 3) with frugivores without harvesting, plants can invade even if they are completely clustered, and 4) with frugivores with harvesting plants can also invade if the local plant density is very high. A complete analysis and mathematical formulars are provided in *Appendix C*.

## 4 Discussion

Plant-frugivore mutualistic interactions are essential for sustaining many economically and socially important plant species (Egerer et al. 2018). However, such mutualistic interactions are increasingly being threatened by anthropogenic activities (Markl et al. 2012). This could potentially alter plant abundances, spatial distribution, and global diversity; and hence, can tremendously affect ecosystem functioning and services. By developing and analyzing a spatially-explicit individual-based model of plant-frugivore-human interactions, we have elucidated the effects of human harvesting of fruits on the persistence, spatial distribution, and establishment of plants, and the ecologically important role played by frugivores in mitigating the negative impact of fruit harvesting on plant populations. The individual-based model exhibits very small variability among different simulation runs (*Appendix D*). We have systematically examined the ecological dynamics of the individual-based model using the method of pair approximation. The pair-approximation model allowed us to analytically study plants establishment and derive useful mathematical formulas describing parameter-dependent conditions under which plants can invade from low abundances in the absence and presence of fruit harvesting. In most cases, we found perfect matching between the results obtained from the individual-based model and those obtained from the pair-approximation model. Under strong fluctuations in the trajectories of the individual-based model (mostly near the plant extinction boundaries), we found very unsignificant mismatching between the two models.

Approximately 90% of tropical plants rely on frugivorous animals for long-distance dispersal of their seeds to maintain their persistence and diversity. Our findings agree with previous theoretical and empirical studies of seed dispersal (Archer and Pyke 1991, Beckman and Rogers 2013, Mohammed et al. 2018), confirming the importance of plant-frugivore interactions for plant persistence and plant spatial distributions. In particular, efficient frugivore species allow plant persistence and gradually shift the spatial distribution of plants from clustered to random through global dispersal of seeds. We have shown how these benefits plants receive from their frugivores depend on the characteristics of plants and their frugivores, especially, the fruit-production rate, the attack rate and seed-dispersal efficiency of frugivores.

Our study has mainly focused on assessing the negative impact of human harvesting of fruits on plant populations supported by frugivore species through provision of seed-dispersal services. We have demonstrated that fruit harvesting can significantly reduce the natural and ecologically important roles played by frugivores for the persistence, spatial distribution, and establishment of plants. Specifically, frugivores increase the global density of plant populations through the global dispersal of seeds and reduce the degree of plant clustering. In contrast, fruit harvesting decreases global plant density and increases plant aggregation, potentially leading to plant extinction. Highly efficient frugivores allow plants to persist in conditions under which they otherwise could not persist, even when harvesting is intense. Fruit harvesting can drive plants to extinction in the absence of frugivores or in the presence of inefficient frugivores, even when plant fecundity is high. Highly efficient frugivores have high potential of reducing the effect of fruit harvesting on global plant density and plant clustering by dispersing seeds globally over the habitat, decreasing local competition and allowing new birth of plants. The intensity of fruit harvesting largely determines the extent to which plants are affected by harvesting, without and with the presence of frugivores. Our results therefore suggest the measurement of fruit-harvesting rates, monitoring frugivore efficiency and plant performance (under the pressure of fruit harvesting) as indicators to protect plant species from harvest- and frugivore-induced extinctions.

Interestingly, our analytical investigations demonstrated that under fruit harvesting plants can not establish from low abundances without frugivores, while frugivores enable plant establishment precluded by fruit harvesting. Efficient frugivore species favour unconditional establishment of plants while inefficient frugivores preclude plant establishment by behaving as seed predators (akin to human harvesting). Significant amount of seeds has to be eaten and deposited for potential successful germination to allow plant to establish from low abundances.

In conclusion, we have developed and analysed an individual-based model of plant-frugivore-human interactions and have elucidated the negative effects of human harvesting of fruits on plant persistence, plant spatial distributions and plants establishment. We have also elucidated the ecologically important role played by frugivore species for mitigating the impact of fruit harvesting on plant populations, particularly enabling plants establishment precluded by harvesting. We have found that fruit harvesting strongly decreases global plant density through two impact routes, by reducing the number of seeds that can produce new plants and by changing the spatial distribution of plants from random to aggregated, thereby aggravating the local competition among plants. While the former impact route is widely acknowledged, our elucidation of the latter impact route will hopefully help protect the ecosystem services rendered by seed-dispersing natural frugivores and thereby promote a more sustainable management of plant-frugivore-human interactions.

## Acknowledgement

We thank Bernd Blasius and Cang Hui for discussion.

## Funding

MM Acknowledges Fellowship from the National Research Foundation of South Africa to participate to the Young Scientists Summer Program at IIASA, the Add-on Fellowship for Interdisciplinary Life Sciences from Joachim Herz Foundation, and the German National Research Foundation in the Research Unit DynaCom (project number 2716).

## Author contribution

MM developed the individual-based and pair-approximation models with significant support from ÅB, UD, and PL; MM performed the numerical simulation, MM prepared the draft; ÅB, UD, and PL commented on, and approved, the draft manuscript.

## Competing interests

The authors declare no competing interests.

## Supplementary information

*Appendix A: Modelling technique using the pair-approximation method*

To complement the stochastic simulations of our individual-based model, we devise and analyze deterministic approximations that enable greater tractability. For this purpose, we briefly recall the methodology of pair approximation we use to approximate the behavior of our individual-based model of plant-frugivore-human interactions. Pair approximation is a method for constructing a system of ordinary differential equations for both global and local densities of a given population (Matsuda et al. 1992, Harada & Iwasa 1994) and allows to formulate spatial dynamics in an analytic fashion (e.g., Hui & Li 2004, Hui & Richardson 2017, Hui et al. 2018). In pair approximation, if the local density is greater than the global density, individuals are aggregated (clustered) in space, equal local and global densities define spatially random structures, and if the local density is lower than the global density, individuals are segregated (overdispersed) in space (Hui et al. 2006).

We first consider an infinitely large regular lattice for the plant population dynamics without animal-mediated seed dispersal and without any external effect such as fruit harvesting. Each lattice site is either occupied (+ site) by an individual plant or empty (0 site). It is assumed that the size of a lattice site only allows one individual plant to establish, that is, each site can only be occupied by a single individual plant at a time. Seeds produced by a plant in a site can only grow in the nearest-neighbouring sites, and only if these are empty. We denote by $P_+$ the probability, called the global density of plants, that a randomly chosen lattice site is occupied by an individual plant, and by $P_0 = 1 - P_+$ the probability that a randomly chosen lattice site is empty. Since the birth of a new plant is restricted by the availability of a vacant nearest-neighboring site, we denote by $q_{0|+}$ the conditional probability that the nearest-neighbouring site of a given occupied site is an empty site, where a seed can germinate and grow into a new plant. By definition, $q_{+|+} = 1 - q_{0|+}$, where $q_{+|+}$ is the conditional probability that the nearest-neighbouring site of a given occupied site is also an occupied site. The probability $q_{+|+}$ is called the local density of plants. According to the pair-approximation methos, the global plant density dynamics is governed by the following equation,

$$\dot{P}_+ = -d_P P_+ + b_P q_{0|+} P_+ = -d_P P_+ + b_P(1 - q_{+|+})P_+, \quad (A1)$$

where $b_P$ and $d_P$ are the natural birth and death rates of plants, respectively. In Eq. 1, the first term refers to the natural death of plants, while the second term refers to the establishment of a new plant in an empty site from locally dispersed seeds of a neighbouring parental plant.

By definition, $q_{+|+} = P_{++}/P_+$, where $P_{++}$ is the probability that two randomly chosen neighbouring sites (a pair) are both occupied. Considering the time derivative (indicated by a dot) gives $\dot{q}_{+|+} = -P_{++}\dot{P}_+/P_+^2 + \dot{P}_{++}/P_+$. To obtain the dynamics $\dot{q}_{+|+}$ of the local density, the pair dynamics $\dot{P}_{++}$ thus needs to be specified. This is given by

$$\dot{P}_{++} = -2d_P P_{++} + 2\frac{b_P}{z}P_{+0} + 2\frac{b_P}{z}(z-1)q_{+|0+}P_{+0}, \tag{A2}$$

where the parameter $z$ represents, as before, the number of a site's nearest-neighboring sites, and measures the local dispersal ability of plants to their neighboring sites. The first term in Eq. 2 describes the transition of a $(+,+)$ pair to either a $(+,0)$ or a $(0,+)$ pair; thus the factor 2. In the second term, an occupied site contributes by the birth of an individual to its nearest-neighboring empty site ($P_{+0} = P_{0+} = q_{0|+}P_+ = (1-q_{+|+})P_+$). In the third term, the empty site in a $(+,0)$ pair may have up to $(z-1)$ occupied neighbouring sites; thus the factor $(z-1)$. Given that $P_{+0} = P_{0+} = (1-q_{+|+})P_+$ and $q_{+|0} = \frac{P_{0+}}{P_0} \leq 1$, we obtain $(1-q_{+|+})P_+ \leq P_0 = (1-P_+)$ which leads to the inequality $2 - 1/P_+ \leq q_{+|+}$. The region under the curve defined by this inequality is mathematically infeasible (see Fig. 4).

The dynamics of the local density of plants can thus be described as a function of only the global density $P_+$ and the local density $q_{+|+}$,

$$\dot{q}_{+|+} = -q_{+|+}[d_P + b_P(1 - q_{+|+})] + \left[\frac{2b_P}{z}\left(1 + (z-1)\frac{(1-q_{+|+})P_+}{(1-P_+)}\right)\right](1 - q_{+|+}). \tag{A3}$$

More details about the derivation of these equations can be found in Matsuda et al. (1992) and Harada & Iwasa (1994).

*Appendix B: Derivation of the local plant density equation*

Here we derive an equation for the local plant density $q_{+|+}$. Given the definition of local plant density $q_{+|+} = \frac{P_{++}}{P_+}$ (see the main text), then the time derivative is given as

$$\dot{q}_{+|+} = -\frac{P_{++}}{P_+^2}\dot{P}_+ + \frac{1}{P_+}\dot{P}_{++},$$

where $P_{++}$ is the probability that two randomly chosen neighboring sites are both occupied. The rate of change of the global plant density $\dot{P}_+$ is described in the main text, and we now need to find an equation for the global plant density of pairs $P_{++}$ (Harada & Iwasa 1994), which is given as

$$\dot{P}_{++} = -2dP_{++} + 2\frac{gr}{z}\frac{s}{s+c+h}P_{+0} + 2\frac{gr}{z}(z-1)\frac{s}{s+c+h}q_{+|0+}P_{+0}$$
$$+ 2gr\mu\frac{c}{s+c+h}P_+P_{+0} - \frac{2hr}{s+c+h}P_{++}.$$

Where z is the number of the nearest-neighboring sites. The first term indicates the transition of a (+,+) pair to (+,0) pair or (0,+) pair, that is where the factor 2 comes from. In the second and third terms, we refer to the birth of the non-dispersed seeds. In the second term, an occupied site contributes by a birth of an individual to its nearest-neighboring empty site with transition from (+,0) pair to (+,+) pair or from (0,+) pair to (+,+) pair. The third term, the presence of an occupied site adjacent to the empty site of given nearest-neighboring sites (+,0) may affect the transition of (+,0) to (+,+), that is, the transition from (+,0,+) to (+,+,+) or (0 → +) could be from any of the neighbors of the 0 site. This is why we multiply by $q_{+|0+}$. The pair-approximation method neglects the effects of the neighbor-of-the neighbor, therefore $q_{+|0+} \approx q_{+|0}$. We have

$$P_{+0} = P_{0+} = P_+ q_{0|+} = P_+(1 - q_{+|+}).$$

$$q_{+|0+} \approx q_{+|0} = \frac{P_{0+}}{P_0} = \frac{(1 - q_{+|+})P_+}{(1 - P_+)}.$$

The $P_{++}$ equation can be written as

$$\dot{P}_{++} = -2dP_{++} + 2\frac{gr}{z}\frac{s}{s+c+h}P_+(1 - q_{+|+}) + 2\frac{gr}{z}(z-1)\frac{s}{s+c+h}\frac{(1 - q_{+|+})^2 P_+^2}{(1 - P_+)}$$
$$+ 2gr\mu\frac{c}{s+c+h}P_+^2(1 - q_{+|+}) - \frac{2hr}{s+c+h}P_{++}.$$

Using the two equations $\dot{P}_+$ and $\dot{P}_{++}$ and the definition provided above, we can now derive an equation to describe the local plant density as follows:

$$\dot{q}_{+|+} = -\frac{P_{++}}{P_+^2}\dot{P}_+ + \frac{1}{P_+}\dot{P}_{++}.$$

$$-\frac{P_{++}}{P_+^2}\dot{P}_+ = \left(d + \frac{hr}{s+c+h}\right)q_{+|+} - gr\frac{s}{s+c+h}(1 - q_{+|+})q_{+|+}$$
$$- g\mu r\frac{c}{s+c+h}q_{+|+}(1 - P_+)\frac{1}{P_+}\dot{P}_{++}$$

$$\frac{1}{P_+}\dot{P}_{++} = -2\left(d + \frac{hr}{s+c+h}\right)q_{++} + 2\frac{gr}{z}\frac{s}{s+c+h}(1 - q_{+|+})$$
$$= -2\left(d + \frac{hr}{s+c+h}\right)q_{++} + 2\frac{gr}{z}\frac{s}{s+c+h}(1 - q_{+|+})$$
$$+ 2\frac{gr}{z}(z-1)\frac{s}{s+c+h}\frac{(1 - q_{+|+})^2 P_+}{(1 - P_+)} + 2gr\mu\frac{c}{s+c+h}P_+(1 - q_{+|+}).$$

$$+ 2\frac{gr}{z}(z-1)\frac{s}{s+c+h}\frac{(1 - q_{+|+})^2 P_+}{(1 - P_+)} + 2gr\mu\frac{c}{s+c+h}P_+(1 - q_{+|+}).$$

Adding these equations, we get

$$\dot{q}_{++} = -\left(d + \frac{hr}{s+c+h}\right)q_{++}$$
$$+ \frac{gr}{z}\frac{s}{s+c+h}\left(2 - z\,q_{+|+} + 2(z-1)\frac{(1-q_{+|+})P_+}{(1-P_+)}\right)(1 - q_{+|+})$$
$$+ g\mu r \frac{c}{s+c+h}(2P_+ - q_{+|+} - P_+ q_{++}).$$

*Appendix C: Plant establishment and plant equilibrium sensitivity analysis*

Using the pair-approximation model, we derive analytical conditions under which plants can invade from low abundances and under which establishment is not possible. We analytically investigate the sensitivity of the plant equilibrium to the model parameters. To analyse plant establishment, we assume that animals are their carrying capacity ($A = K$) while the global plant density, $P_+$, is set to zero. However, the local plant density $q_{+|+}$, defined by pair-approximation as $P_{++}/P_+$, has a non-zero value when the global plant density is zero. We first recall the global plant density equation describing the global dynamics of plants as

$$\dot{P}_+ = -d_P P_+ + gr(1 - q_{+|+})\frac{s}{c+s+h}P_+ + \mu gr(1 - P_+)\frac{c}{c+s+h}P_+ - \frac{h}{c+s+h}rP_+. \quad (C1)$$

We denote by $G$ the per-capita growth rate at $P_+ = 0$ and $A = K$

$$G = -d_P + gr(1 - q^*_{+|+})\frac{s}{aK+s+h} + \mu gr\frac{aK}{aK+s+h} - \frac{h}{aK+s+h}r. \quad (C2)$$

By setting $G > 0$ we derive parameter-dependent conditions under which plants can invade. Rearranging equation (2) we get

$$(1 - q^*_{+|+}) > (1/grs)[d_P(aK + s + h) - \mu graK + hr]. \quad (C3)$$

Where $(1 - q^*_{+|+}) = q^*_{0|+}$, the probability of finding empty local site.

This means $q^*_{0|+}$ should be large, $q^*_{+|+}$ small, that is, space has to be uncrowded for succesful establishment

$$q^*_{+|+} < 1 + \left(\frac{1}{grs}\right)[\mu graK - d_P(aK + s + h) - hr]. \quad (C4)$$

We take the derivatives of the right-hand side ($R$) of equation (C4) and see how it changes with respect to parameters. Let

$$R = 1 + \left(\frac{1}{grs}\right)[\mu graK - d_P(aK + s + h) - hr].$$

The derivative with respect to the harvesting rate $h$

$$\frac{dR}{dh} = -(d_P + r)/(grs) < 0.$$

That means that $R$ decreases as $h$ increases, indicating that the inequality (eqn. C4) may not hold for large $h$. In this case plants can not invade for large $h$. It is ecologically obvious that plants with low abundances can not persist under the pressure of fruit harvesting.

The derivative with respect to the natural death rate $d_P$

$$\frac{dR}{dd_P} = -(h + s + Ka)/(grs) < 0.$$

That means that $R$ decreases as $d_P$ increases, indicating that the inequality (eqn. C4) may not hold for large $d_P$. In this case plants can not invade for large $d_P$. The high mortality rate can easily exceed the birth rate of plants with low abundances, causing establishment from low abundance to be impossible.

The derivative with respect to the fruit-production rate $r$

$$\frac{dR}{dr} = \frac{d_P(h + s + Ka)}{gsr^2} > 0.$$

That means that $R$ increases as $r$ increases, indicating that the inequality (eqn. C4) may hold for large $r$. In this case plants can invade for large $r$. When seeds is highly abundant, then plants can invade from low abundance.

The derivative with respect to the natural dispersal rate $s$

$$\frac{dR}{ds} = ((d_P h + hr + Kad_P) - Kag\mu r)/(grs^2).$$

That means that $R$ increases as $s$ increases ($\frac{dR}{ds} > 0$) only if

$$(1/Ka)(d_P h + hr + Kad_P) > g\mu r,$$

hence natural dispersal favours establishment under this condition. For large $d_P$ or large $h$ this inequality does not hold and thus plants may not invade from low abundance for large $s$.

The derivative with respect to the germination probability $g$

$$\frac{dR}{dg} = \frac{d_P h + d_P s + hr + Kad_P}{g^2 rs} > 0.$$

That means that $R$ decreases as $g$ increases, indicating that the inequality (eqn. C4) may hold for large $g$, hence high germination probability favours establishment.

The derivative with respect to the carrying capacity of animals $K$

$$\frac{dR}{dK} = (ag\mu r - ad_P)/(grs).$$

That means that $R$ increases as $K$ increases ($\frac{dR}{dK} > 0$) only if $g\mu r > d_P$, hence animal's carrying capacity favours establishment under this condition. This establishment condition indicates that the rate of new birth of plants due to animal dispersal must exceed the death rate of plants independent of the new birth of plants due to local dispersal.

The derivative with respect to the animal attack rate $a$

$$\frac{dR}{da} = (Kg\mu r - Kd_P)/(grs).$$

That means that $R$ increases as $a$ increases ($\frac{dR}{da} > 0$) only if $g\mu r > d_P$, hence animal attack rate favours establishment under this condition. This establishment condition indicates that the rate of new birth of plants due to animal dispersal must exceed the death rate of plants independent of the new birth of plants due to local dispersal.

The derivative with respect to the dispersal efficiency $\mu$

$$\frac{dR}{d\mu} = \frac{Ka}{s} > 0.$$

That means that $R$ increases as $\mu$ increases, hence animal dispersal efficiency favours establishment.

The global plant density equilibrium for the system without frugivores is

$$P_+^* = -(d_P h + d_P s + hr - grs)/(grs).$$

An analytical solution for the system equations with frugivores does unfortunately not exist.

The derivative with respect to the natural death rate $d_P$ is

$$\frac{dP_+^*}{dd_P} = -(h + s)/(grs) < 0.$$

The derivative with respect to $h$ is

$$\frac{dP_+^*}{dh} = -(dP + r)/(grs) < 0.$$

The derivative with respect to the seed-production rate $r$ is

$$\frac{dP_+^*}{dr} = (d_P(h + s))/(gr^2 s) > 0.$$

The derivative with respect to the natural dispersal rate $s$ is

$$\frac{dP_+^*}{ds} = (h(dP + r))/(grs^2) > 0.$$

The derivative with respect to the germination probability $g$ is

$$\frac{dP_+^*}{dg} = (d_P h + d_P s + hr)/(g^2 rs) > 0.$$

## *Appendix D: The effect of other model parameters on the plant-frugivore-human system*

Here we demonstrate the dependence on the other model parameters of the plant-frugivore-human system. The results presented below (Fig. 9-15) are quite intuitive and expected.

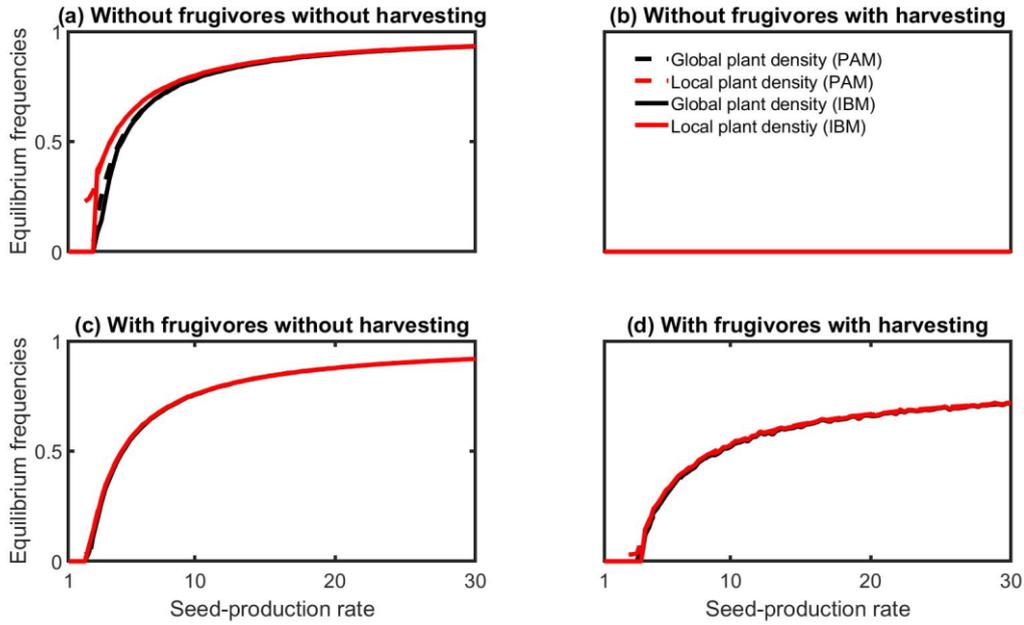

**Fig. 9:** Changes in the equilibrium frequencies of global plant density $P_+$ (continuous and dashed black curves) and local plant density $q_{+|+}$ (continuous and dashed red curves) -- predicted by the individual-based model (continuous curves) and the pair-approximation model (dashed curves) -- for different fruit-production rates: (a) without frugivores and without fruit harvesting, (b) without frugivores and with fruit harvesting (c) with frugivores and without fruit harvesting, and (d) with frugivores and with fruit harvesting. All other parameter values are as indicated in the caption of Figure 2.

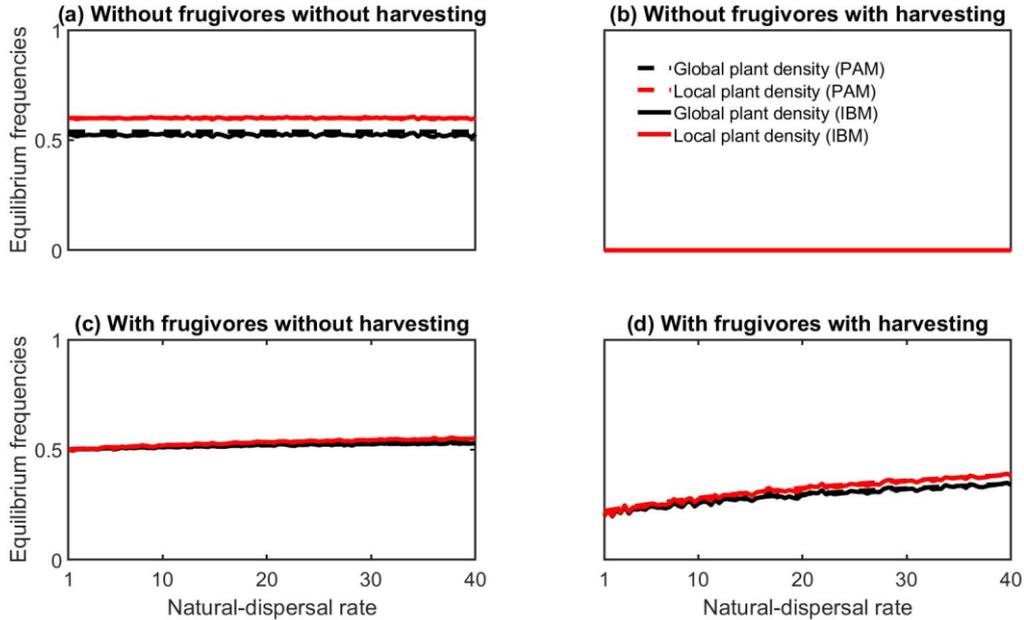

**Fig. 10:** Changes in the equilibrium frequencies of global plant density $P_+$ (continuous and dashed black curves) and local plant density $q_{+|+}$ (continuous and dashed red curves) with natural dispersal rate of plants -- predicted by the individual-based model (continuous curves) and the pair-approximation model (dashed curves) -- for four ecological scenarios: (a) without frugivores and without fruit harvesting, (b) without frugivores and with fruit harvesting (c) with frugivores and without fruit harvesting, and (d) with frugivores and with fruit harvesting. All other parameter values are as indicated in the caption of Figure 2.

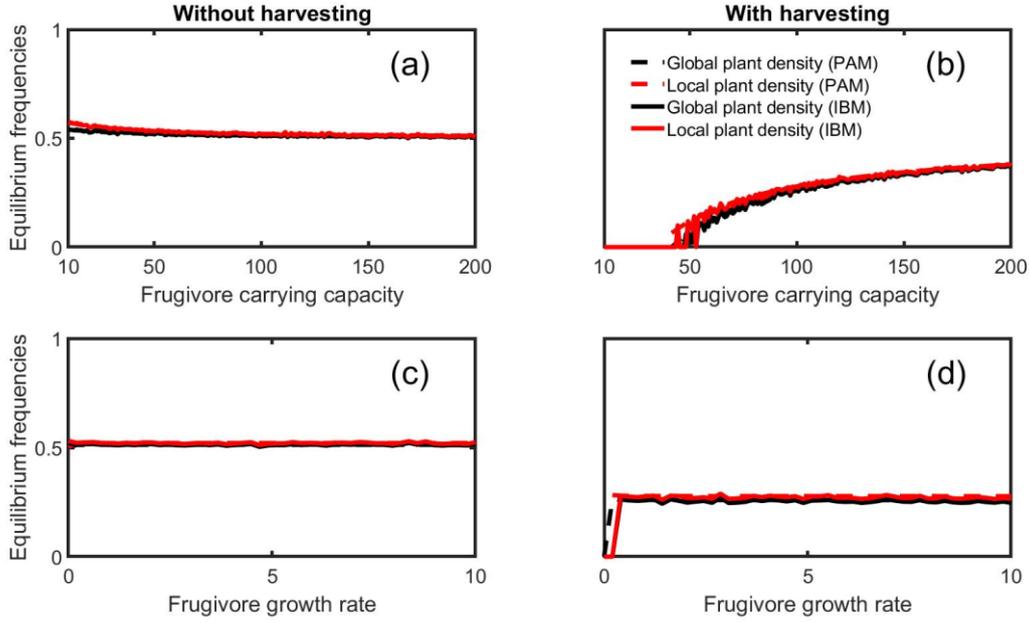

**Fig. 11:** Changes in the equilibrium frequencies of global plant density $P_+$ (continuous and dashed black curves) and local plant density $q_{+|+}$ (continuous and dashed red curves) with the carrying capacity of frugivores (panels a and b) and frugivore growth rate (panel c and d) -- predicted by the individual-based model (continuous curves) and the pair-approximation model (dashed curves) -- without (first column) and with (second column) fruit harvesting. All other parameter values are as indicated in the caption of Figure 2.

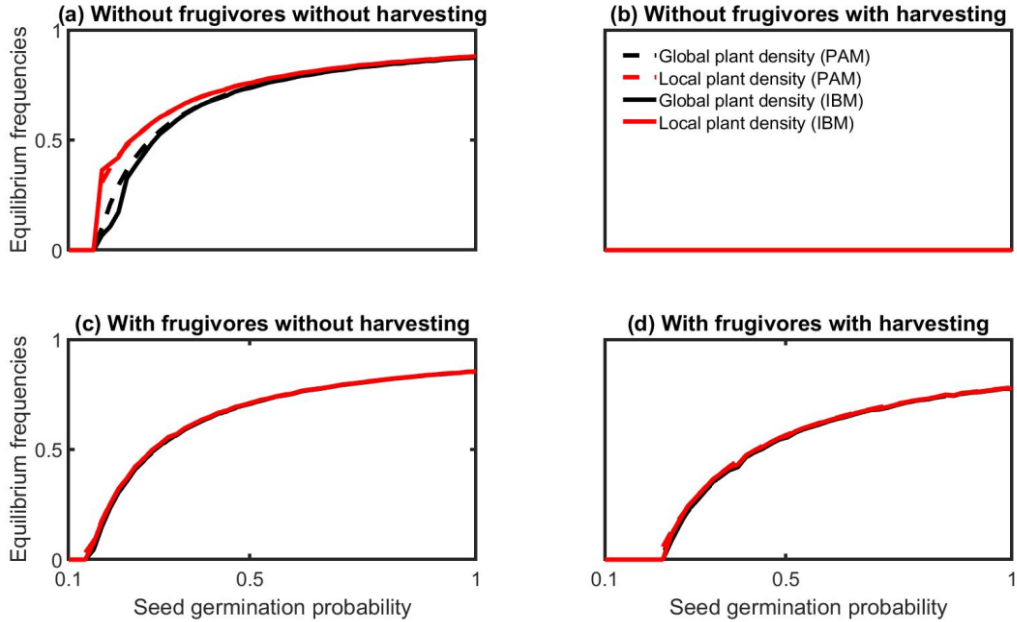

**Fig. 12:** Changes in the equilibrium frequencies of global plant density $P_+$ (continuous and dashed black curves) and local plant density $q_{+|+}$ (continuous and dashed red curves) with seed germination probability -- predicted by the individual-based model (continuous curves) and the pair-approximation model (dashed curves) -- for four ecological scenarios: (a) without frugivores and without fruit harvesting, (b) without frugivores and with fruit harvesting (c) with frugivores and without fruit harvesting, and (d) with frugivores and with fruit harvesting. All other parameter values are as indicated in the caption of Figure 2.

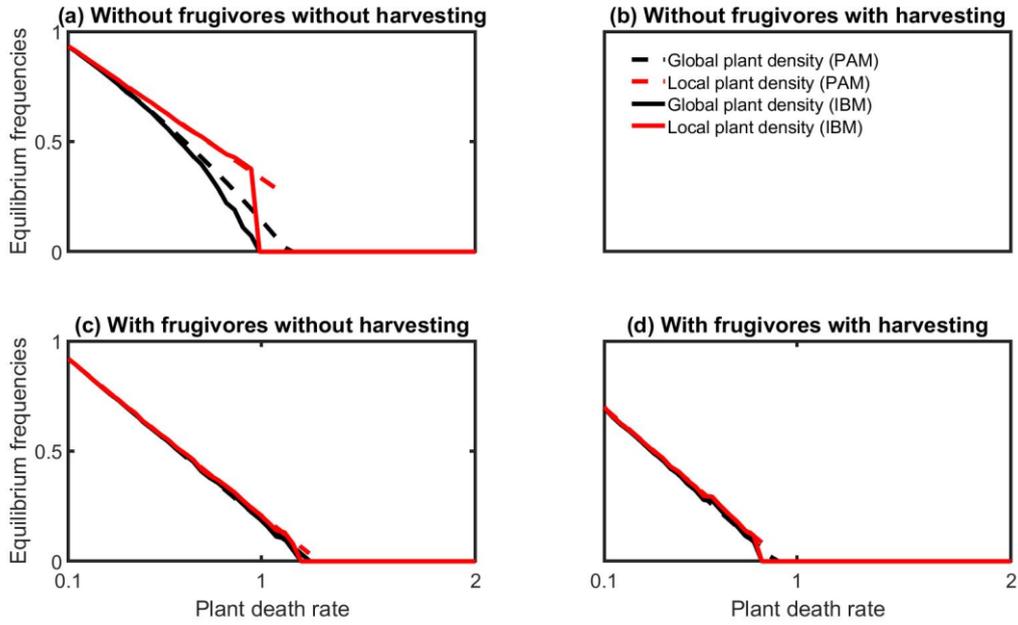

**Fig. 13:** Changes in the equilibrium frequencies of global plant density $P_+$ (continuous and dashed black curves) and local plant density $q_{+|+}$ (continuous and dashed red curves) with natural death rates of plants -- predicted by the individual-based model (continuous curves) and the pair-approximation model (dashed curves) -- for four ecological scenarios: (a) without frugivores and without fruit harvesting, (b) without frugivores and with fruit harvesting (c) with frugivores and without fruit harvesting, and (d) with frugivores and with fruit harvesting. All other parameter values are as indicated in the caption of Figure 2.

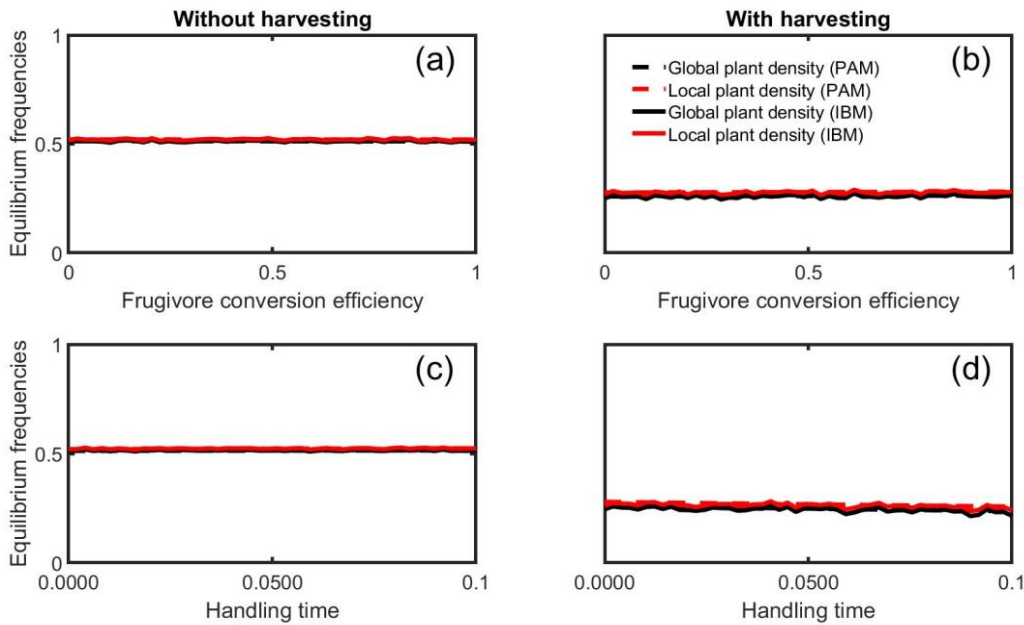

**Fig. 14:** Changes in the equilibrium frequencies of global plant density $P_+$ (continuous and dashed black curves) and local plant density $q_{+|+}$ (continuous and dashed red curves) with frugivore conversion efficiency (panel a and b) and handling time (panel c and d) -- predicted by the individual-based model (continuous curves) and the pair-approximation model (dashed curves) without and with harvesting. All other parameter values are as indicated in the caption of Figure 2.

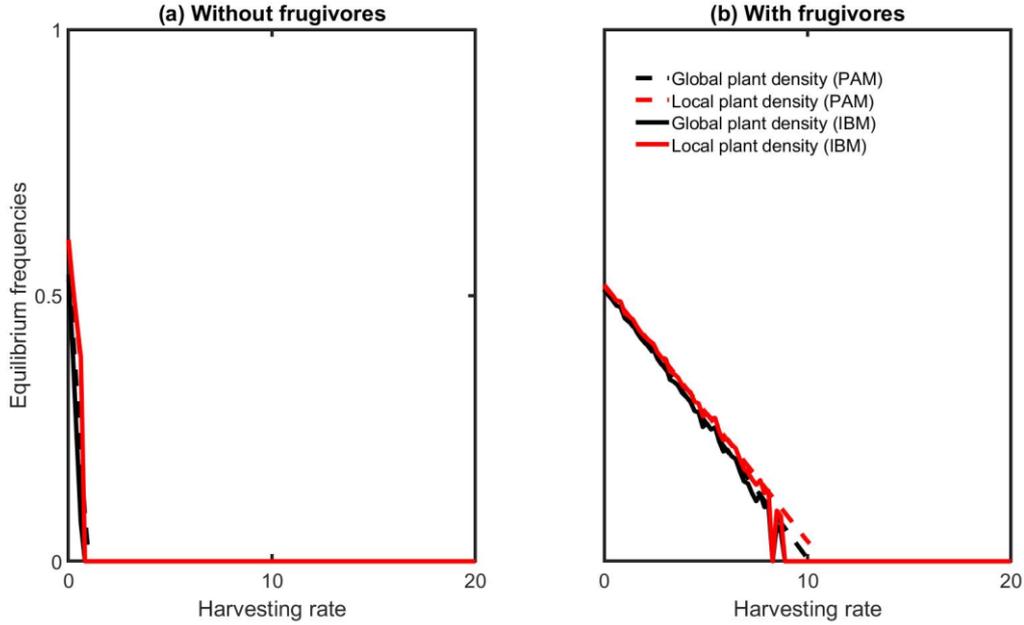

**Fig. 15:** Changes in the equilibrium frequencies of global plant density $P_+$ (continuous and dashed black curves) and local plant density $q_{+|+}$ (continuous and dashed red curves) with fruit harvesting rates -- predicted by the individual-based model (continuous curves) and the pair-approximation model (dashed curves) – for two different ecological scenarios: (a) without frugivores, and (b) with frugivores. All other parameter values are as indicated in the caption of Figure 2.

## *Appendix E: Comparisons between IBM, PAM, and mean-field approximation model*

The following figures show good agreement between the different models, and Figure 18 in particular shows the variability of the statistical equilibrium from the IBM simulations is almost zero.

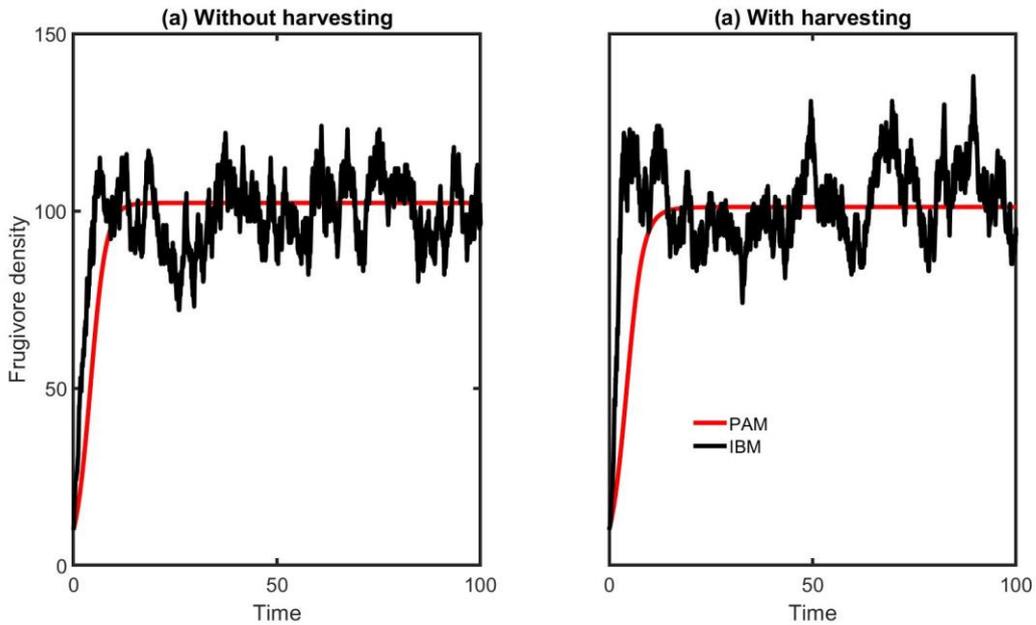

**Fig. 16:** The effect on fruit harvesting on density of frugivores. The black curve obtained from the IBM simulation and the red curve obtained from the PAM simulation. All other parameter values are as indicated in the caption of Figure 2.

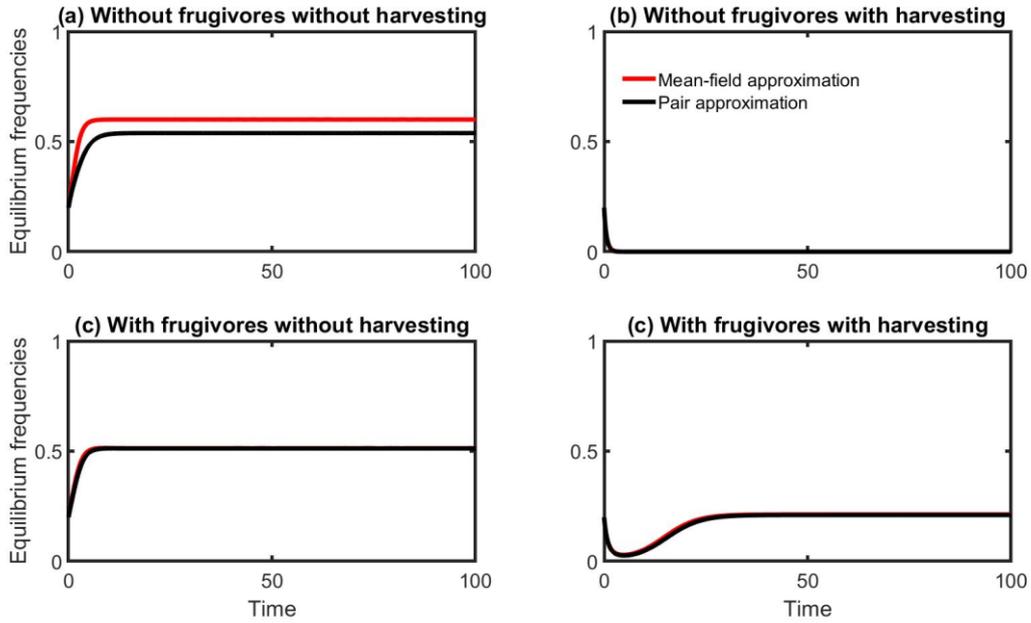

**Fig. 17:** Changes in the frequencies of global plant density $P_+$ over time -- predicted by mean-field approximation (red curves) and pair approximation (black curves) – for four ecological scenarios: (a) without frugivores and without fruit harvesting, (b) without frugivores and with fruit harvesting (c) with frugivores and without fruit harvesting, and (d) with frugivores and with fruit harvesting. All other parameter values are as indicated in the caption of Figure 2.

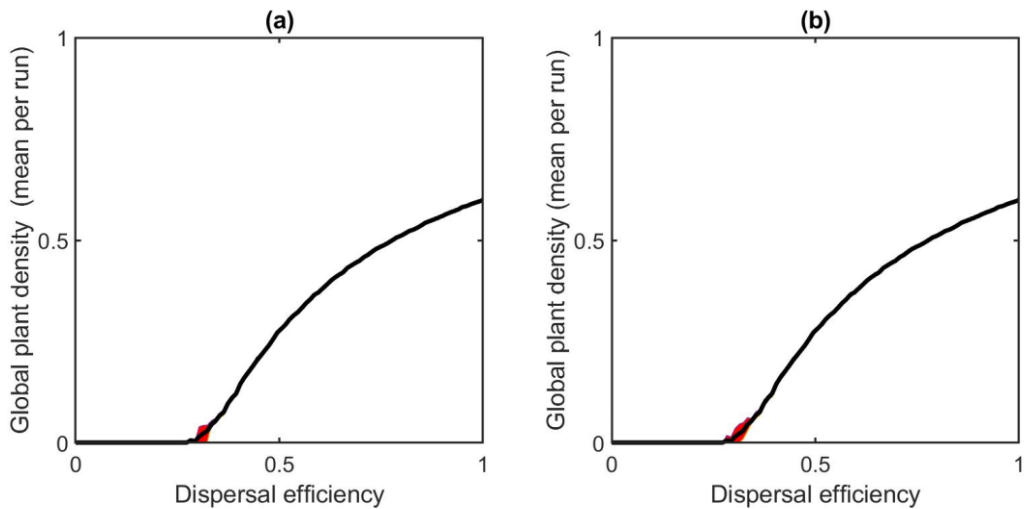

**Fig. 18:** Small variability among 10 simulation runs of our individual-based model. The black curves represent the mean global plant density as a function of dispersal efficiency and averaged over 10 stochastic realizations. The black shaded-region show (a) one unit of standard deviation below and above the mean, and (b) the lower and upper quantiles. All other parameter values are as indicated in the caption of Figure 2.